\documentclass[singlespacing]{elsart}

\usepackage{graphicx}

\usepackage{amssymb}
\journal{}
\begin{document}

\begin{frontmatter}

\title{Probability distribution function for reorientations in Maier-Saupe potential}

\author{A.E. Sitnitsky},
\ead{sitnitsky@kibb.knc.ru}

\address{Kazan Institute of Biochemistry and Biophysics, P.O.B. 30, Kazan
420111, Russia. e-mail: sitnitsky@kibb.knc.ru }

\begin{abstract}
Exact analytic solution for the probability distribution function of the non-inertial rotational diffusion equation, i.e., of the Smoluchowski one, in a symmetric Maier-Saupe uniaxial potential of mean torque is obtained via the confluent Heun's function. Both the ordinary Maier-Saupe potential and the double-well one with variable barrier width are considered. Thus, the present article substantially extends the scope of the potentials amenable to the treatment by reducing Smoluchowski equation to the confluent Heun's one. The solution is uniformly valid for any barrier height. We use it for the calculation of the mean first passage time. Also the higher eigenvalues for the relaxation decay modes in the case of ordinary Maier-Saupe potential are calculated. The results obtained are in full agreement with those of the approach developed by Coffey, Kalmykov, D\'ejardin and their coauthors in the whole range of barrier heights.
\end{abstract}

\end{frontmatter}

\newpage
\section{Introduction}
Rotational reorientations take place in liquid crystals, polymers, proteins, lipids and many other systems. They are investigated by means of variety of experimental techniques such as NMR, dielectric relaxation spectroscopy, fluorescence depolarization, etc (see, e.g., \cite{Bri96} and refs. therein). The basic tool for their theoretical description is the Fokker-Planck equation [2]-[5]. For the case when the inertial effects are negligible (overdamped limit) the latter is reduced to the Smoluchowski equation (SE). There is vast literature on their study and applications [2]-[46]. The applications of them to the dielectric spectroscopy {\cite{Mar71}, \cite{Sto85}, \cite{Cof06}, fluorescence depolarization \cite{Zan83} and NMR relaxation of liquid crystals (see \cite{Don97}, \cite{Don02}, \cite{Don10} and refs. therein) have been thoroughly explored. Also, the extension of the theory from ordinary diffusion to the fractional one is intensively studied (see \cite{Cof04}, and refs. therein). Besides SE arises in the theory of ferromagnetism (where it is called Brown's equation) \cite{Bro63}, \cite{Bro79}, \cite{Cof12}. Both in the theory of molecular rotational motion in a uniaxial potential \cite{Dej97}, \cite{Kal09}, \cite{Cof04} (occurring for dynamics of liquid crystals, dielectric relaxation, etc.) and in the theory of ferromagnetism \cite{Cof12} the so called Maier-Saupe potential is widely used and by right considered to be a basic model.

The most powerful approach to the problem was developed by Coffey, Kalmykov, D\'ejardin and their coauthors in the above mentioned papers \cite{Cof01}, \cite{Cof84}, \cite{Cof04}, \cite{Cof06}, \cite{Kal09}, \cite{Kal11}. For the sake of brevity we further call it as CKD. The approach is based on the expansion of the probability distribution function as a series of spherical harmonics. This method is well suited for the potentials of mean torque that can be expanded in terms of spherical harmonics. It results in an infinite hierarchy of differential-recurrence relations for the moments (the expectation values of the spherical harmonics). On the one hand CKD is an exact approach that uses no approximations and imposes no physical limitations. The undoubted merit of CKD is that it is very general and can be applied to enormous variety of different physical situations described by corresponding types of potentials. On the other hand this approach makes use of foreign eigenfunctions (Legendre polynomials) for SE under consideration. For this reason, a complementary approach was suggested by the author \cite{Sit15} that yields an exact solution in purely mathematical sense. The latter uses the expansion of the solution over own eigenfunctions of the Smoluchowski operator rather than that over Legendre polynomials. The own eigenfunctions are expressed via the confluent Heun's function (CHF). The advantage of our approach is that it enables one to avoid dealing with infinite hierarchies of differential-recurrence relations that is a complex and laborious problem in itself. Besides it yields directly the probability distribution function itself rather than statistical moments (i.e., values averaged over it). Such option is convenient for, e.g., applications of the theory to relaxation processes in NMR.

However the approach of \cite{Sit15} was developed for a rather specific potential (double-well Maier-Saupe one with fixed barrier width parameter). The main aim of the present paper is to show that our approach has wider applicability. We substantially extend the range of potentials amenable to treatment by this method and apply it to both the ordinary and the double-well Maier-Saupe potentials while the latter may have variable barrier width parameter. The consideration is carried out within the framework of a uniform approach based on the general Maier-Saupe potential of which both ordinary and double-well options are merely particular cases. Nevertheless, despite the above mentioned extension of the options our heuristic potential is still a symmetric uniaxial Maier-Saupe mean field one
\[
 \frac{V(\psi)}{k_BT}=\sigma \sin^2 \psi-2c\ln \sin \psi
\]
where $\sigma$ is the barrier height parameter and $c$ is the barrier width parameter that takes the values from the range $0 \leq c < \infty$. The variable barrier width parameter $c$ enables one to eliminate the logarithmic term at wish. The standard uniaxial potential
\[
\frac{V(\psi)}{k_BT}=\sigma \sin^2 \psi
\]
is merely a particular case $c=0$ of the previous one so that both cases are amenable to exact analytic treatment via the CHF in a uniform manner. The particular case $c=0$ was explored in details by CKD in \cite{Dej97}, \cite{Kal09}, \cite{Cof04}. The results obtained there are the reliable reference data for the present approach. The SE for the standard form of the potential $c=0$ can be as well solved exactly in terms of known functions in the same way as that for the potential with the logarithmic term $c>0$. From the purely mathematical point of view it seems natural to treat both cases simultaneously. The particular case with fixed barrier width parameter $c=1$ explored in \cite{Sit15} is distinguished by extreme simplicity of the boundary conditions for the characteristic equation for the spectrum of eigenvalues.

From the physical point of view the potential with the logarithmic term is considered for the following reason.
For rigid rotators without internal motions (a nematic liquid crystal molecule or a fine single-domain ferromagnetic particle) the minima of a uniaxial potential over the polar angle $\psi$ in the laboratory frame are at $\psi=0$ and $\psi=\pi$. In these cases the standard Maier-Saupe potential is relevant and there are no reasons to expect that the mean field theory can provide the appearance of the logarithmic term in such systems. In contrast the latter finds visible physical motivation within the context of internal (segmental) motions in polymers and proteins (e.g., those of $\Omega$-loops). The minima of a uniaxial potential over the polar angle $\psi$ in the molecular frame for a  segment of structure are not necessarily at $\psi=0$ and $\psi=\pi$ and we need a genuinely double-well potential to describe such motions. The logarithmic term in the potential is a simplest way to shift the equilibrium positions from $\psi=0$ and $\psi=\pi$. However an important difference should be stressed. The segmental motions can not proceed with the isotropy of the potential over the azimuthal angle. In most cases they proceed with a fixed azimuthal angle that requires a different normalization of the probability distribution function than for the case of isotropy (see \cite{Sit15} for more details). Otherwise the consideration follows a similar line as for the case of isotropy. In the present article we choose the normalization with the isotropy of the potential over the azimuthal angle for the sole reason: to provide the possibility of direct comparison of our results with those of the approach developed by CKD. We show that our approach is valid for this case and hence can be sure that it will remain valid for a realistic case of the potential with a fixed azimuthal angle.

As mentioned above our approach yields directly the probability distribution function for SE. The latter enables one to obtain any required average in a uniform manner. We calculate with its help the mean first passage time to demonstrate that our approach leads to identical results with those of CKD. We pay special attention to the case of ordinary Maier-Saupe potential because for it there is a full set of data from CKD. In our opinion it is of interest to compare the results of the present approach based on reducing SE to confluent Heun's equation with those of CKD for this case. On the one hand as mentioned above the ordinary Maier-Saupe potential is a basic theoretical model for materials based on both nematic liquid crystals and ferromagnetics. On the other hand the case of ordinary uniaxial Maier-Saupe potential despite of its simplicity is the cornerstone basic example for the theory. Hence, any discrepancies in this key issue are impermissible. We show that there is no  divergence between the results of two approaches in this case. There is an important issue where our approach can be especially helpful. In \cite{Cof951}, \cite{Cof941} the problem of higher eigenvalues for the relaxation decay modes was raised. It is related to the question whether the relaxation proceeds via a single mechanism or not. Within the framework of CKD the calculation of higher eigenvalues requires laborious numerical work by the method developed in \cite{Cof941}. In our approach the eigenvalues are the roots of the a characteristic equation for the corresponding boundary problem for the Smoluchowski operator. This equation includes CHF and its derivative. Maple makes finding its roots a routine procedure in all practically important cases. However, some technical problems arise at calculating higher eigenvalues (e.g., $\lambda_5$) by Maple for small barriers. They presumably result from known deficiencies of realization of CHF derivative  in this symbolic computational software package. We discuss the issue below.

The paper is organized as follows.  In Sec. 2 the problem under study is formulated.  In Sec. 3 the solution of SE is presented. In Sec. 4 the spectrum of eigenvalues is derived. In Sec. 5 the probability distribution function is obtained. In Sec.6 the MFPT is calculated with its help. In Sec. 7 the results are discussed and the conclusions are summarized.

\section{Smoluchowski equation}
For a uniaxial potential of mean torque $V(\psi)$ SE under consideration is \cite{Dej97}, \cite{Kal09}, \cite{Cof04}
\[
2\tau\frac{\partial f(\psi,t)}{\partial t}=\frac{\beta}{\sin \psi}
\frac{\partial}{\partial \psi}\left[\sin \psi f(\psi,t)\frac{\partial V(\psi)}{\partial \psi}\right]+
\]
\begin{equation}
\label{eq1}
\frac{1}{\sin \psi}\frac{\partial}{\partial \psi}\left[\sin \psi \frac{\partial f(\psi,t)}{\partial \psi}\right]
\end{equation}
where $\tau$ is the characteristic relaxation time for isotropic non-inertial rotational diffusion (e.g., $\tau_D$ the Debye one for the theory of dielectric relaxation or $\tau_N$ the N\'eel one for the theory of ferromagnetism) and $\beta=1/\left(k_BT\right)$. Its solution is the probability distribution function $f(\psi, t)$ that must be normalized so that its integral over the whole space gives
\begin{equation}
\label{eq2} \frac{1}{4\pi}\int \limits_{0}^{2\pi} d\varphi
\ \int \limits_{0}^{\pi} d\psi
\ \sin \psi \ f(\psi, t)=\frac{1}{2}\int \limits_{0}^{\pi} d\psi
\ \sin \psi \ f(\psi, t)=1
\end{equation}
The normalized initial condition takes the form
\begin{equation}
\label{eq3} f(\psi, 0)=\frac{2}{\sin \psi_0}\ \delta (\psi-\psi_0)
\end{equation}

We introduce a new variable
\begin{equation}
\label{eq4} x=\cos\psi
\end{equation}
Then the equation takes the form
\begin{equation}
\label{eq5} \frac{\partial f(x,t)}{\partial t}=\frac{1}{2\tau}
\frac{\partial}{\partial x}\left[(1-x^2)\left(\frac{\partial f(x,t)}{\partial x}+\beta f(x,t)V'(x)\right)\right]
\end{equation}
where the dash means the derivative over variable $x$.

Further we consider a general Maier-Saupe uniaxial potential of mean torque potential
\begin{equation}
\label{eq6} V(x)=\frac{\sigma}{\beta}\left(1-x^2\right)-\frac{c}{\beta}\ln (1-x^2)
\end{equation}
The ordinary Maier-Saupe uniaxial potential of mean torque potential considered in \cite{Dej97}, \cite{Kal09}, \cite{Cof04} corresponds to the particular case $c=0$. The double-well Maier-Saupe potential with fixed barrier width parameter suggested in \cite{Sit15} corresponds to the particular case $c=1$. Both barrier height $E$ and width $w$ depend on $\sigma$ and $c$
\[
E=V\left(x_{max}\right)-V\left(x_{min}\right)=\frac{\sigma}{\beta}\left(1-\frac{c}{\sigma}\right)+
\frac{c}{\beta}\ ln\left(\frac{c}{\sigma}\right)
\]
\begin{equation}
\label{eq7} w=x_{min}^{(2)}-x_{min}^{(1)}=2\sqrt{1-\frac{c}{\sigma}}
\end{equation}
Inversely one obtains
\[
\sigma=\beta E\left\{\frac{w^2}{4}+\left(1-\frac{w^2}{4}\right)ln\left(1-\frac{w^2}{4}\right)\right\}^{-1}
\]
\begin{equation}
\label{eq8} c=\beta E \left(1-\frac{w^2}{4}\right)
\left\{\frac{w^2}{4}+\left(1-\frac{w^2}{4}\right)ln\left(1-\frac{w^2}{4}\right)\right\}^{-1}
\end{equation}
For the ordinary Maier-Saupe potential ($c=0$) we have $w=2$. The dependence of the potential (\ref{eq6}) on the parameters is depicted in Fig.1. The requirement for the potential to be a double-well one is $c<\sigma$.

\begin{figure}
\begin{center}
\includegraphics* [width=\textwidth] {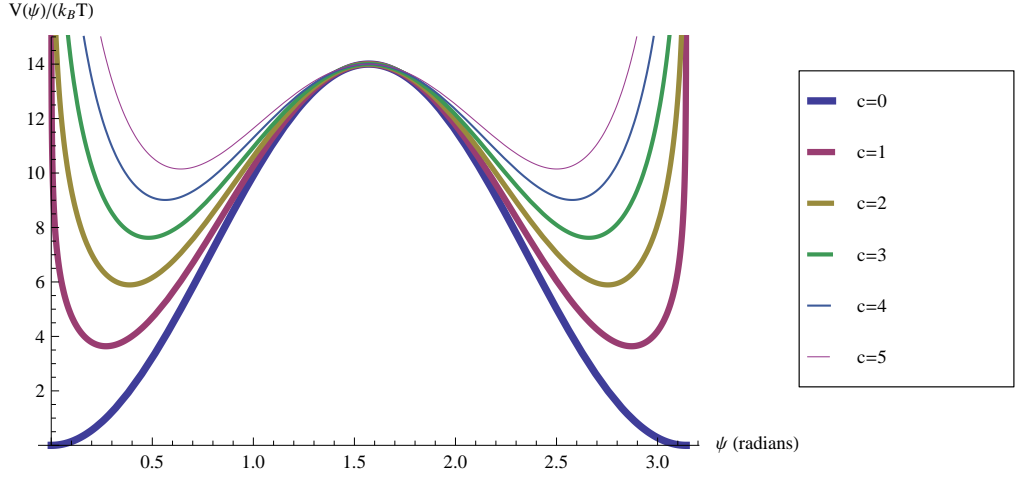}
\end{center}
\caption{The symmetric Maier-Saupe uniaxial potential of mean torque $V(\psi)/\left(k_B T\right)=\sigma\ \sin^2 \psi-c\ ln \left (\sin^2 \psi\right)$. For $c>0$ it is called the double-well Maier-Saupe potential to distinguish it to the ordinary Maier-Saupe one $c=0$. The value of the parameter $\sigma$ is $\sigma=14$.} \label{Fig.1}
\end{figure}

\section{Solution of Smoluchowski equation}
Equation (\ref{eq5}) can be solved by separation of variables
\begin{equation}
\label{eq9} f(x, t)=\eta(x)\exp\left(-\rho t\right)
\end{equation}
The equation for the function $\eta(x)$ is
\[
\left(1-x^2\right)\eta''_{xx}(x)-2x\left[\sigma \left(1-x^2\right)-(c-1)\right]\eta'_{x}(x)+
\]
\begin{equation}
\label{eq10} 2\left[\rho \tau+\left(c-\sigma\right)+3\sigma x^2\right]\eta (x)=0
\end{equation}
We introduce a new variable
\begin{equation}
\label{eq11} y=x^2
\end{equation}
The equation takes the form
\[
y(y-1)\eta''_{yy}(y)+\left[-\sigma y^2+\frac{1}{2}\left(1+2\sigma-2(c-1)\right)y-\frac{1}{2}\right]\eta'_{y}(y)-
\]
\begin{equation}
\label{eq12}
\frac{1}{2}\left[\rho \tau+\left(c-\sigma\right)+3\sigma y\right]\eta(y)=0
\end{equation}
It belongs to a class of the so-called confluent Heun's equation \cite{Ron95}. Equation (\ref{eq12}) has fundamental solutions that can be expressed via the confluent Heun's function (CHF). The latter is a known special function \cite{Ron95}, \cite{Fiz12}, \cite{Fiz10}. At present it is realized explicitly in the only symbolic computational software package Maple as $HeunC$ and its derivative $HeunCPrime$ (see \cite{Fiz12} for detailed discussion of the merits and drawbacks of this computational tool in Maple). Returning to the variable $x$ we can write the general solution of (\ref{eq10}) as follows
\begin{equation}
\label{eq13}\eta(x)=C\eta^{(1)}(x)+D\eta^{(2)}(x)
\end{equation}
where
\begin{equation}
\label{eq14}\eta^{(1)}(x)=HeunC^{(1)}\left(-\sigma,-\frac{1}{2},-c,-\frac{\sigma}{2}\left(c+\frac{3}{2}\right),
\frac{\sigma/2-\rho \tau}{2}-\frac{c-1}{4};x^2\right)
\end{equation}
\begin{equation}
\label{eq15}\eta^{(2)}(x)= xHeunC^{(2)}\left(-\sigma,\frac{1}{2},-c,-\frac{\sigma}{2}\left(c+\frac{3}{2}\right),\frac{\sigma/2-\rho \tau}{2}-\frac{c-1}{4};x^2\right)
\end{equation}

\section{Spectrum of eigenvalues}
The boundary conditions are obtained from (\ref{eq10}) by setting $x=\pm1$
\begin{equation}
\label{eq16} 2x(c-1)\eta'_{x}(x)+ 2\left[\rho \tau+c+2\sigma \right]\eta (x)=0
\left| {\begin{array}{l}
  \\
x=\pm1\\
 \end{array}}\right.
\end{equation}
The characteristic equation for our boundary problem is split apart to a pair of equations (omitting for the sake of brevity the parameters of CHFs and their derivatives)
\begin{equation}
\label{eq17}2(c-1)HeunCPrime^{(1)}\left(;1\right)+(\rho \tau+c+2\sigma)HeunC^{(1)}\left(;1\right)=0
\end{equation}
\begin{equation}
\label{eq18}2(c-1)HeunCPrime^{(2)}\left(;1\right)+(\rho \tau+2c-1+2\sigma)HeunC^{(2)}\left(;1\right)=0
\end{equation}
In case $c=1$ they are reduced to simpler ones \cite{Sit15}
\begin{equation}
\label{eq19}HeunC^{(1)}\left(;1\right)=0
\end{equation}
\begin{equation}
\label{eq20}HeunC^{(2)}\left(;1\right)=0
\end{equation}
We denote
\begin{equation}
\label{eq21} \Delta_n=\rho_n^{(1)} \tau
\end{equation}
\begin{equation}
\label{eq22} \Omega_m=\rho_m^{(2)} \tau
\end{equation}
The equations (\ref{eq17}) and (\ref{eq18}) can be solved only numerically but Maple copes with this problem. One always obtains $\Delta_1=0$ while the lowest $\Omega_1$ is always nonzero and the corresponding term mainly determines the behavior of the distribution function $f(\psi,t)$ and MFPT calculated with its help. The behavior of $\Omega_1$ is depicted in Fig.2.
\begin{figure}
\begin{center}
\includegraphics* [width=\textwidth] {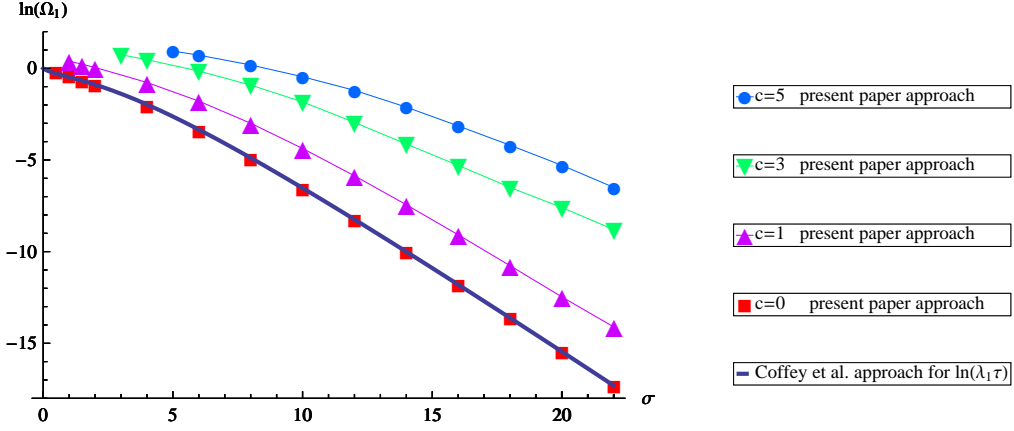}
\end{center}
\caption{The dependence of the first non-zero eigenvalue for the boundary problem of rotational Smoluchowski equation on the parameter $\sigma$. The latter defines to a greater extent the barrier height of the double-well Maier-Saupe potential. The continuous line is the result of the approach developed by Coffey, Kalmykov, D\'ejardin and their coauthors (\ref{eq32}) for the particular case of ordinary Maier-Saupe potential.}
\label{Fig.2}
\end{figure}
Higher eigenvalues are plotted in Fig.4 and Fig.5.

\section{Probability distribution function}
We return to the variable $\psi$ and write the solution of SE (\ref{eq1}) as
\[
f(\psi,t)= C_1\eta^{(1)}_1(\psi)+D_1 \exp \left(-\Omega_1 t/\tau\right)\eta^{(2)}_1(\psi)+
\]
\begin{equation}
\label{eq23}C_2\exp \left(-\Delta_2 t/\tau\right)\eta^{(1)}_2(\psi)+D_2 \exp \left(-\Omega_2 t/\tau\right)\eta^{(2)}_2(\psi)+...
\end{equation}
The coefficients $C_n$ and $D_m$ are calculated with taking into account the fact that being the solution of the boundary problems $\eta^{(1)}_n(\psi)$ and $\eta^{(2)}_m(\psi)$ are orthogonal of each other
\begin{equation}
\label{eq24} \int \limits_{0}^{\pi} d\psi
\ \sin \psi\ \eta^{(k)}_n(\psi)\eta^{(l)}_m(\psi)=\delta_{mn}\delta_{kl}
\end{equation}
where  $k,l=1,2$ and $n,m=0,1, 2, ...\ $.
At time $t=0$  (\ref{eq23}) yields (with taking into account (\ref{eq3}))
\begin{equation}
\label{eq25}\frac{2\delta (\psi-\psi_0)}{\sin \psi_0}= C_1 \eta^{(1)}_1(\psi)+
D_1 \eta^{(2)}_1(\psi)+C_2\eta^{(1)}_2(\psi)+D_2 \eta^{(2)}_2(\psi)+...
\end{equation}
Then the coefficients $C_n$ and $D_m$ are obtained by multiplying (\ref{eq25}) by $\eta^{(1)}_n(\psi)$ or $\eta^{(2)}_m(\psi)$ and integration over $\psi$ with taking into account (\ref{eq24})
\begin{equation}
\label{eq26} C_n=2\eta^{(1)}_n(\psi_0)\left[\int \limits_{0}^{\pi} d\psi
\ \sin \psi\ \left[\eta^{(1)}_n(\psi)\right]^2\right]^{-1}
\end{equation}
and
\begin{equation}
\label{eq27} D_m=2\eta^{(2)}_m(\psi_0)\left[\int \limits_{0}^{\pi} d\psi
\ \sin \psi\ \left[\eta^{(2)}_m(\psi)\right]^2\right]^{-1}
\end{equation}
Thus, we have the explicit algorithm for obtaining the coefficients $C_n$ and $D_m$ along with the corresponding eigenvalues $\Delta_n$ and $\Omega_m$ that makes the solution of our problem to be completed. Substitution of all these values into (\ref{eq23}) yields the required probability distribution function.

\section{Mean first passage time}
To exhibit how the result obtained may be useful we calculate with its help the dependence of MFPT on the parameters of the potential $\sigma$ and $c$. Our effective Maier-Saupe uniaxial potential of mean torque is
\begin{equation}
\label{eq28} \frac{1}{k_B T}V(\psi)=\sigma\ \sin^2 \psi-c\ ln \left (\sin^2 \psi\right)
\end{equation}
It has two local minima the left of which we denote as $\psi_L$ and the maximum at $\psi_M=\pi/2$. Following Risken \cite{Ris89} one can obtain for the non-averaged MFPT $T_1$ that characterizes the ability for the system to reach the barrier top $\psi_M$ starting from the left minimum $\psi_0=\psi_L$ the expression (see, e.g., \cite{Sit15} for details)
\begin{equation}
\label{eq29} T_1 (\psi_0)=-\int \limits_{0}^{\psi_M} d\psi\ \sin \psi
\int \limits_{0}^{\infty} dT\ T\ \frac{\partial f(\psi, T)}{\partial T}
\end{equation}
After straightforward calculations we obtain
\begin{equation}
\label{eq30} \frac{1}{\tau}T_1 (\psi_0)=\int \limits_{0}^{\psi_M} d\psi\ \sin \psi\ \left\{ \frac{D_1}{\Omega_1}\eta^{(2)}_1(\psi)+
\frac{C_2}{\Delta_2}\eta^{(1)}_2(\psi)+\frac{D_2}{\Omega_2}\eta^{(2)}_2(\psi)+...\right\}
\end{equation}
Finally, we calculate the averaged MFPT $<T_1>$ for the case when the initial condition is thermodynamically averaged over the left well
\[
<T_1>=\left [\int \limits_{0}^{\pi} d\psi_0\ \sin \psi_0 \exp\left(-\frac{V( \psi_0)}{k_BT}\right)\right]^{-1}
\times
\]
\begin{equation}
\label{eq31}\int \limits_{0}^{\psi_M} d\psi_0\ \sin \psi_0 \exp\left(-\frac{V( \psi_0)}{k_BT}\right)T_1 (\psi_0)
\end{equation}
It is worthy to recall here that the coefficients $C_n$ and $D_m$ are the function of the initial condition $\psi_0$. The results of numerical calculation for the averaged MFPT $<T_1>$ are presented in Fig.3.
\begin{figure}
\begin{center}
\includegraphics* [width=\textwidth] {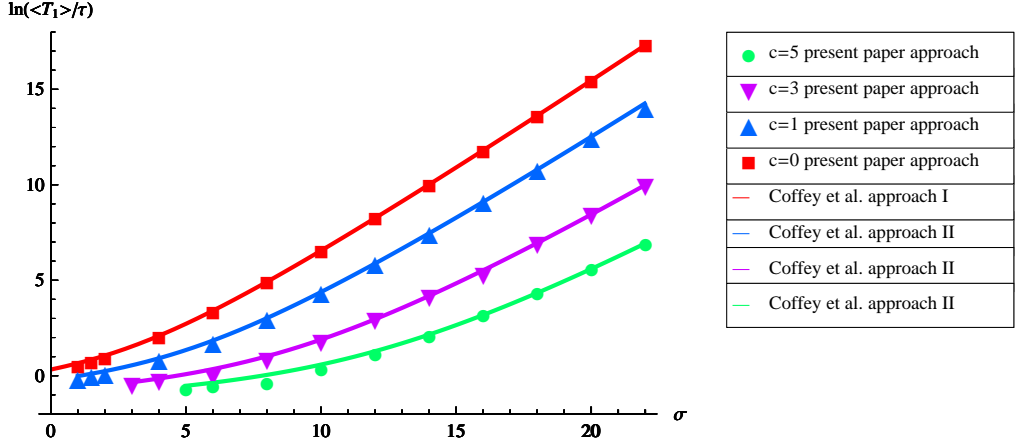}
\end{center}
\caption{The dependence of the mean first passage time for the transition from the well of the symmetric double-well Maier-Saupe uniaxial potential of mean torque $V(\psi)/\left(k_B T\right)=\sigma\ \sin^2 \psi-c\ ln \left (\sin^2 \psi\right)$ on the parameter $\sigma$. The latter defines to a greater extent the barrier height of the double-well Maier-Saupe potential. The continuous line for the particular case of ordinary Maier-Saupe potential ($c=0$) is the result of the full-fledged (time dependent) approach of Coffey, Kalmykov, D\'ejardin and their coauthors (\ref{eq33}). The continuous lines for cases with $c>0$ are the result of the restricted approach of Coffey, Kalmykov, D\'ejardin and their coauthors (\ref{eq34}) within the framework of the stationary approximation.}
\label{Fig.3}
\end{figure}

\section{Results and discussion}
In Fig.2 the first non-zero eigenvalue for the ordinary Maier-Saupe potential ($c=0$) is compared with the result of CKD for $\lambda_1\tau$ given by the empirical equation \cite{Cof04}, \cite{Cof12} which is valid for all barrier heights
\begin{equation}
\label{eq32}\lambda_1\tau=\frac{\sigma}{e^{\sigma}-1}\left[2^{-\sigma}+
\frac{2\sigma^{3/2}}{\sqrt{\pi}\left(1+\sigma\right)}\right]
\end{equation}
Our results are in excellent agreement with those of CKD for the case of ordinary Maier-Saupe potential ($c=0$).

In Fig.3 our results for MFPT for the ordinary Maier-Saupe potential ($c=0$) are compared with those of CKD \cite{Cof04}, \cite{Cof12}
\[
 \frac{<T_1>}{\tau}=4\Biggl \{ \left[\int \limits_{-1}^{0} dz\ \frac{e^{-\sigma z^2}}{1-z^2}\int \limits_{-1}^{z} dz'\ e^{\sigma z'^2}\right]^{-1}+
\]
\begin{equation}
\label{eq33}\left[\int \limits_{0}^{1} dz\ \frac{e^{-\sigma z^2}}{1-z^2}\int \limits_{-1}^{z} dz'\ e^{\sigma z'^2}\right]^{-1}\Biggr \}^{-1}
\end{equation}
We call this formula CKD I. It is obtained within the framework of a full-fledged (time dependent) approach taking into account the transient dynamics. One can see that our approach is in excellent agreement with CKD I in the whole range of barrier heights. The points in Fig. 3 are calculated with the only mode $\Omega_1$ everywhere but for the values $\sigma=1;1.5;2$. For the latter we have to add the modes $\Omega_2$ and $\Delta_2$ to provide agreement with (\ref{eq33}).

There are no results of full-fledged CKD for double-well Maier-Saupe potential ($c>0$) in the literature. For this reason in Fig. 3 we compare our results for this case with those of restricted CKD obtained in the stationary approximation. The latter in our case of the potential (\ref{eq28}) is eq. (1.18.1.7) of \cite{Cof04}
\[
 \frac{<T_1>}{\tau}=2 \int \limits_{0}^{\pi/2} d\psi'\ \frac{e^{\sigma\ \sin^2 \psi'-c\ ln \left (\sin^2 \psi'\right)}}{\sin \psi'}\times
\]
\begin{equation}
\label{eq34}\int \limits_{0}^{\psi'} d\psi\ \sin \psi\ e^{-\sigma\ \sin^2 \psi+c\ ln \left (\sin^2 \psi\right)}
\end{equation}
We call it CKD II. In this case the results of two approaches are also in quite satisfactory agreement for the whole range where the potential remains to be a double-well one ($c<\sigma$). Outside this range the very notion of MFPT loses its meaning. As earlier all the points in Fig. 3 for the cases $c>0$ are calculated with the only mode $\Omega_1$ everywhere but for corresponding lowest values of $\sigma$. For the latter we have to add the modes $\Omega_2$ and $\Delta_2$ to provide agreement with (\ref{eq34}).

Fig. 4 shows that higher eigenvalues are also in excellent agreement with those of CKD. The values of $\lambda_3$ and $\lambda_5$ for comparison with our results are taken from Table.1 of \cite{Cof941}. However, the implementation of computations within the framework of our approach is much more easier than that by the method of \cite{Cof941}. It is accomplished with the help of Maple practically automatically. Nevertheless, Maple has problems to find the roots for $\Omega_m$ with $m\geq3$ below some critical barrier heights, e.g., those for $\lambda_5$ (our $\Omega_3$) at the values of $\sigma=1$, $\sigma=2$ and $\sigma=3$ (see Fig.4). The same phenomenon takes place for $\Delta_n$ with $n\geq3$ as Fig.5 shows. We stress that none of these missed roots is necessary for the calculation of MFPT in the agreement with CKD. As was noted above there at most $\Omega_2$ and $\Delta_2$ are invoked to at some marginally low barrier heights. For the particular case $c=1$ when the characteristic equation does not include $HeunCPrime$ (see (\ref{eq19}) and (\ref{eq20})) there was no problem with calculation of eigenvalues at any $\sigma$ \cite{Sit15}. Thus, we have drawn the conclusion that the phenomenon may be a result of known deficiencies of realization of $HeunCPrime$ in Maple (see \cite{Fiz12} for expert opinion on the drawbacks of this computational tool). We hope that future development of numerical realizations of this special function in symbolic computational software packages will make its usage in the problem under consideration more convenient and reliable.
\begin{figure}
\begin{center}
\includegraphics* [width=\textwidth] {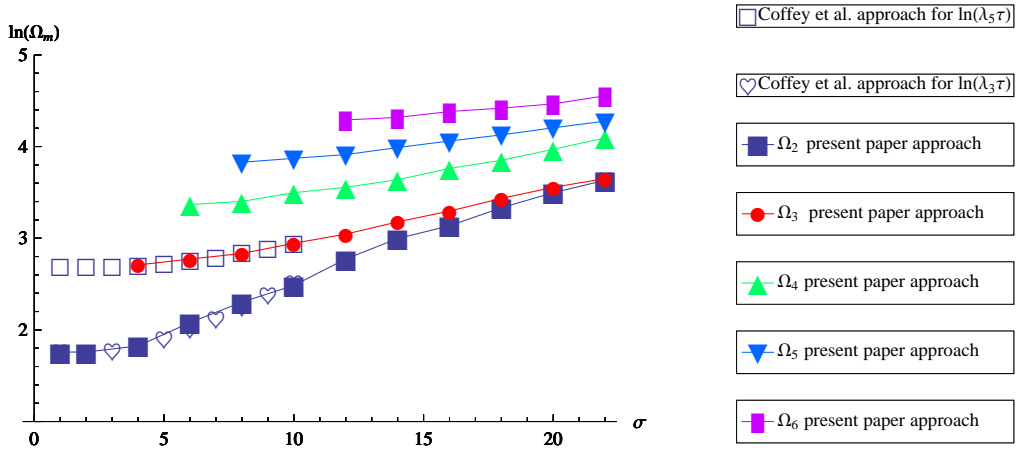}
\end{center}
\caption{The dependence of higher eigenvalues $\Omega_m$ ($\tau \lambda_{odd}$ in the approach of Coffey, Kalmykov, D\'ejardin and their coauthors) on the parameter $\sigma$ for the ordinary Maier-Saupe potential (the particular case $c=0$ of the general potential (\ref{eq6})). The parameter $\sigma$ defines the barrier height of the potential.}
\label{Fig.4}
\end{figure}
\begin{figure}
\begin{center}
\includegraphics* [width=\textwidth] {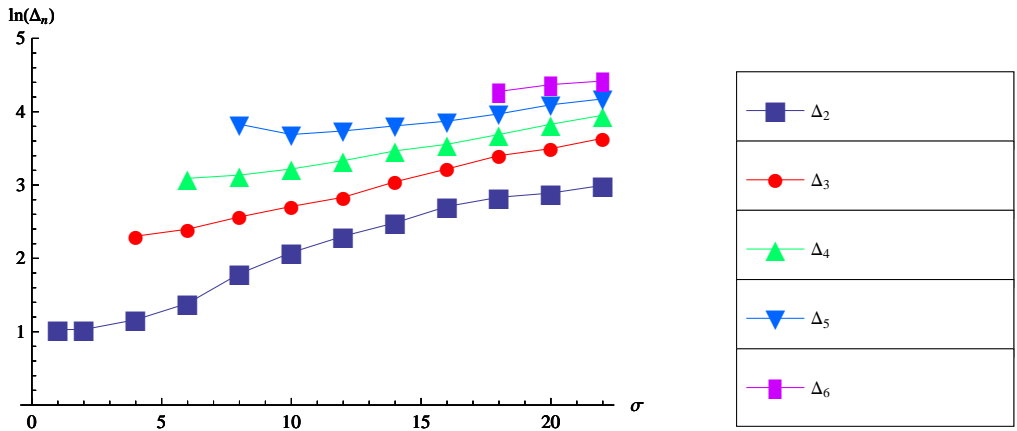}
\end{center}
\caption{The dependence of higher eigenvalues $\Delta_n$ on the parameter $\sigma$ on the parameter $\sigma$ for the ordinary Maier-Saupe potential (the particular case $c=0$ of the general potential (\ref{eq6})). The parameter $\sigma$ defines the barrier height of the potential.}
\label{Fig.4}
\end{figure}

Comparing our approach with CKD we conclude that it has both merits and drawbacks. On the one hand it is conceptually simpler because it makes use of own eigenfunctions for the Smoluchowski operator rather than foreign ones. As a consequence to deal with infinite series of closed expressions for independent coefficients is simpler than with infinite hierarchies of differential-recurrence relations where all coefficients are interdependent of one another. Also our approach provides the option to deal directly with the probability distribution function rather than with statistical moments (i.e., values averaged over it). On the other hand it is applicable only to symmetric Maier-Saupe potential. Besides, CKD provides a user of the theory with compact analytical formulas for MFPT while our approach leads to the expressions that can be treated only numerically.

The approach developed in the present article can not be directly applied to the asymmetric potentials, e.g., to that
\[
 \frac{V(\psi)}{k_BT}=\sigma \sin^2 \psi-\xi \cos \psi
\]
taking place for a uniaxial crystal subjected to a dc bias field. The latter is considered in details by CKD in \cite{Dej97}, \cite{Kal09}, \cite{Cof04}. The reason of failure is purely mathematical: odd powers of $x$ in the potential (we remind that $\cos \psi=x$) hinder the crucial substitution $y=x^2$ (\ref{eq11}) leading to the terms of the type $\sqrt y$ that make the resulting equation to be intractable.
However there are some prospects to generalize the present approach to the potential of the type
\[
 \frac{V(\psi)}{k_BT}=\sigma \sin^2 \psi-\xi F(\cos \psi, \sin \psi) -2c\ln \sin \psi
\]
where $F(\cos \psi, \sin \psi)$ is an arbitrary function and $\xi$ is the asymmetry parameter. In what follows we denote $F(x)\equiv F\left(x,\sqrt{1-x^2}\right)$ and $F'_x$ means the derivative of this function over $x$. The particular case mentioned above is $F(x)=x$; $c=0$. In this case we have SE
\[
\frac{\partial f}{\partial t}=L_Sf+\frac{\xi}{2\tau}\frac{\partial }{\partial x}\left [\left(1-x^2\right)fF'_x\right]
\]
where $L_S$ is the Smoluchowski operator. The solution for the unperturbed case $\xi=0$ obtained in the present article yields its eigenfunctions
\[
L_S\eta^{(1)}_n=-\frac{\Delta_n}{\tau}\eta^{(1)}_n\ \ \ \ \ \ \ \ \ \ \ \ \ \ \ \ L_S\eta^{(2)}_m=-\frac{\Omega_m}{\tau}\eta^{(2)}_m
\]
They are eigenfunctions of the boundary problem and thus form a complete system of orthogonal functions so that any function can be decomposed in a series over them. In particular we have
\[
\frac{\partial }{\partial x}\left [\left(1-x^2\right)\eta^{(1)}_nF'_x\right]=\sum_{k}P^n_k\eta^{(1)}_k+\sum_{l}Q^n_l\eta^{(2)}_l
\]
\[
\frac{\partial }{\partial x}\left [\left(1-x^2\right)\eta^{(2)}_mF'_x\right]=\sum_{i}R^m_i\eta^{(1)}_i+\sum_{j}S^m_j\eta^{(2)}_j
\]
Here $P^n_k$, $Q^n_l$, $R^m_i$ and $S^m_j$ are well defined coefficients, e.g.
\[
P^n_k=-\int \limits_{-1}^{1} dx\ \left(1-x^2\right)\eta^{(1)}_nF'_x\frac{\partial \eta^{(1)}_k}{\partial x}
\]
etc. We seek the solution of SE in the form
\[
f=\sum_{n}A_n(t)\eta^{(1)}_n+\sum_{m}B_m(t)\eta^{(2)}_m
\]
After straightforward calculations with taking into account the redesignation of dumb indices we obtain the system of equations for the coefficients
\[
\frac{dA_n}{dt}=-\frac{\Delta_n}{\tau}A_n+\frac{\xi}{2\tau}\left\{ \sum_{k}A_kP^k_n+\sum_{i}B_iR^i_n \right \}
\]
\[
\frac{dB_m}{dt}=-\frac{\Omega_m}{\tau}B_m+\frac{\xi}{2\tau}\left\{ \sum_{l}A_lQ^l_m+\sum_{j}B_jS^j_m \right \}
\]
This system resembles the infinite hierarchies from CKD in the sense that each coefficient depends on all others. It can be solved by a perturbation method providing that $\xi$ is sufficiently small ($\xi << 1$). However the present article is devoted to the stringent approach to the solution of SE and considering perturbative techniques is beyond its scope.

The general case of the potential $V(\psi, \varphi)$ can not be treated by the approach developed in the present article. The latter works only if the motion over the polar angle $\psi$ is separated somehow from that over the azimuthal one $\varphi$.

In conclusion, we show that the approach based on solving the Smoluchowski equation via the confluent Heun's function has a wide range of applicability to symmetric uniaxial Maier-Saupe potential from the ordinary case to the double-well one with variable barrier width. The solution is uniformly valid for any barrier height. Our results for both mean first passage time and eigenvalues are in good agreement with those of Coffey, Kalmykov, D\'ejardin and their coauthors in the whole range of barrier heights.

Acknowledgements. The author is grateful to Dr. Yu.F. Zuev
for helpful discussions. The work was supported by the grant from RFBR N15-44-02230.

\end{document}